\documentclass[11pt]{article}
  \usepackage{amssymb}
    \usepackage{graphicx}
\newcommand{\sect}[1]{\setcounter{equation}{0}\section{#1}}

\def\am{angular~momentum~}

\def\al {\alpha}
\def\av{angular velocity~}
\def\avv{angular velocities~}

\def\ba{\begin{eqnarray}}
\def\bA{\mbox{\bm$A$\ubm}~}

\def\bB{\mbox{\bm$B\!$\ubm}~}

\def\be{\begin{equation}}

\def\bi{\bibitem}
\def\bJ{\mbox{\bm$J$\ubm}~}
\def\bm{\boldmath}
\def\bn{\mbox{\bm$n\!\!$\ubm}~}
\def\bna{\mbox{\bm$\nabla\!$\ubm}~}

\def\bt {\beta}

\def\B {\overline}

\def\Br{\B r}

\def\cd{\!\cdot}

\def\coo{coordinates~}
\def\cy{cylinder~}

\def\de{\delta}

\def\di{\partial}

\def\ea{\end{eqnarray}} 
\def\ed{electrodynamics~}
\def\ee{\end{equation}}
\def\eee{equation~}
\def\eeee{equations~}
\def\em{energy-momentum~}
\def\ema{electromagnetic~}

\def\ep{\epsilon}

\def\fr{\frac}

 \def\ga{\gamma}
\def\gf{gravitational ~field~}

 \def\gmn{g_{\mu\nu}}

\def\gm{gravomagnetic~}

\def\GR{General Relativity~}

\def\ha{\frac{1}{2}~}
\def\hy{hypersurface~}

\def\inf{\infty}

\def\ka{\kappa}

\def\lb{\label}
\def\les{\ \lesssim}

\def\lll{\left(}

\def\LLL{\left[}

\def\mn{\mu\nu}

\def\na{\nabla}

\def\nn{\nonumber}
\def\nnn{\noindent}

\def\om{\omega}

\def\Om {\Omega}

\def\OO{{\cal O}}

\def\ra{\rightarrow}

\def\rrr{\right)}

\def\Ra{\Rightarrow}
\def\RRR {\right]}

\def\s{{\it one$\!$}~}
\def\si{\sigma}
\def\sq{\sqrt}

\def\sim{\simeq}
\def\sp{spheroid~}
\def\spp{spheroids~}
\def\ss{{\it two$\!$}~}
\def\sss{spacetime~}
\def\ssss{spacetimes~}
\def\st{stationary~}

 \def\Si{\Sigma}
\def\Sc{Schwarzschild~}

\def\td{\tilde}

\def\te{\theta}

\def\tha{{\ts{\ha}\!}}
\def\ti{\times}

\def\tmu{\td\mu}

\def\tr{\td r}

\def\ts{\textstyle}

\def\tte{\td\theta}

\def\Te{\Theta}

\def\ubm{\unboldmath}

\def\vf{\varphi }

\def\vp{\varpi}

\def\vs{\vskip 0.5 cm}

\def\1{{\it one}}
\def\1k{\fr{1}{\ka}}

\def\2k{\fr{1}{2\ka}}

\def\4{\ts{\fr{1}{4}}}

\begin{document}

\title{{\bf Centrifugal force induced by relativistically rotating spheroids and cylinders   }}
\author{Joseph Katz$^{1,2 }$\thanks{email:
jkatz@phys.huji.ac.il} \, Donald Lynden-Bell$^{2}$\thanks{email:dlb@ast.cam.ac.uk}
 \, Ji\v r\'\i ~Bi\v c\'ak$^{2,3}$\thanks{email:bicak@mbox.troja.mff.cuni.cz}
 \\
\\  {\it$^1$ The Racah Institute of Physics, Givat Ram, 91904 Jerusalem, Israel}
\\\\ {\it$^2$  Institute of Astronomy, Madingley Road, Cambridge CB3 0HA,
United Kingdom}\\
\\ {\it$^3$  Institute of Theoretical Physics,  Charles University, 180 00 Prague 8, Czech Republic} \\
{\it and Max Planck Institute of Gravitational Physics, Albert Einstein Institute, }
\\{\it Am M\"uhlenberg 1,
 D-14476 Golm, Germany}}
\maketitle
\begin{abstract} 
\setlength{\baselineskip}{20pt plus2pt}
 Starting from the gravitational potential of a Newtonian spheroidal shell we discuss
electrically charged rotating prolate spheroidal shells in the Maxwell theory. In particular
we consider two confocal charged shells which rotate oppositely in such a way that there
is no magnetic field outside the outer shell. In the Einstein theory
we solve the Ernst equations in the region where the long prolate spheroids are almost
cylindrical; in equatorial regions the exact Lewis "rotating cylindrical" solution is so derived
by a limiting procedure from a spatially bound system.

    In the second part we analyze two cylindrical shells rotating in opposite directions in such a way that the static Levi-Civita metric is produced outside and no angular momentum flux escapes to infinity. The rotation of the local inertial frames in flat space inside the inner cylinder is thus exhibited without any approximation or interpretational difficulties within this model.
    
A test particle within the inner cylinder kept at rest with respect to axes that do not rotate as seen from infinity experiences a centrifugal force. Although the spacetime there is Minkowskian out to the inner cylinder nevertheless that space has been induced to rotate, so relative to the  local inertial frame the particle is traversing a circular orbit. 

\nnn PACS numbers  04.20.-q
\end{abstract}

\setlength{\baselineskip}{20pt plus2pt}
\sect{Introduction}
We aim to give a neat demonstration of centrifugal force on a static body which is induced by the rotation of a heavy shell that surrounds it. The shell causes the Minkowski space inside it to rotate, so, relative to that space, the static body moves backwards on a circle and experiences centrifugal force. 

Pfister and Brown \cite{PB} have earlier studied this problem up to second order in $\Om$ in a distorted sphere, however the problem can be more neatly solved to all orders using rotating cylinders. This was done by Embacher \cite{Em83} generalizing work by Frehland \cite{Fr}. Earlier papers on rotating cylindrical shells where by Papapetrou  {\it et al} \cite{PMS} and Jordan and McCrea \cite{JM}. 

The crucial property of cylinders is that centrifugal force-induced tensions retain the symmetry (unlike those in a sphere).

However the use of infinite rotating cylinders implies that space is not asymptotically flat and since the \sss near the axis is Minkowskian there is some ambiguity in deciding which axes should be considered as non-rotating. We remove these difficulties by showing that the cylindrical \eeee are recovered as an approximation to the spaces inside and in between two tall prolate spheroidal shells which have no net \am as they rotate in opposite senses about their axis. Outside both shells the space is static and tends asymptotically to   flat \Sc space at   infinity. We demonstrate the strong analogy between rotating cylindrical shells in gravitation theory and solenoids in Maxwell's \ed$\!\!$. In the latter the magnetic flux that runs through a solenoid of finite length returns as a magnetic field outside it. As the solenoid is made longer and longer the flux returns over a wider and wider area, so, in the limit, as an infinite solenoid has a field strength of zero magnitude outside which nevertheless carries finite flux. This is the reason why there is no gravomagnetic field outside a rotating infinite cylinder. To ensure that all rotational effects are confined we shall treat two cylinders rotating in opposite directions so as to give no net \am and no gravomagnetic flux outside the outer one.


\sect{Gravomagnetism and electromagnetism} 
There is a strong analogy between \st \ema fields and solutions of \st metrics in \GR$\!\!$. Consider the \st metric
\be
ds^2=\xi^2(dt+A_idx^i)^2 - \ga_{ij}dx^idx^j ~~~i, j=1, 2, 3;~~~\xi=e^{-\psi}.
\lb{21}
\ee
In the Newtonian limit $\psi$ is small and $\bA\!\!=0$. Even in strong field \GR Landau and Lifshitz's  \eee may be rewritten, using 3-space metric's $(\bna\!\!\ti\!\bA\!\!)^j=\eta^{jkl}\di_kA_l$ where $\eta^{jkl}$ is the alternating symbol divided by $\sq{\det(\ga_{ij})}$, in the form
\be\!
\bna\!\ti\!(\xi^3\bB)\!=2\ka\bJ~~~,~~~\bB=(\bna\!\ti\!\bA\!\!)~~~,~~~\ka=\fr{8\pi G}{c^4}.
\lb{22}
\ee
Here the divergensless current
\be
J^i=\xi(T^i_0 - \tha\de^i_0 T)=\1k\xi R^i_0.
\lb{23}
\ee
The above \eeee display the analogy to Maxwell's with magnetic permeability $\xi^{-3}$. We shall therefore quote results of analogous \ema problems to guide our understanding of solutions of Einstein's equations.

In Newtonian gravitation a homoeoidal shell of mass $M$ on a prolate spheroid of semi-axes $a>b$ has the potential
\be
 \psi = m\ts{\ln\lll   \sq{1+\fr{a^2}{\tr^2}}+\fr{a}{\tr}  \rrr}=\ln\lll     \fr{\Br+a}{\Br-a}          \rrr^{m/2}\!\!~,~r\ge b \nn~,~
 \psi =\ts{ \fr{GM}{a}\ln\lll   \sq{1+\fr{a^2}{b^2}}+\fr{a}{b}  \rrr}={\rm const}.~,~r\le b,
\lb{24}
\ee
The last \eee is Newton's theorem; in (\ref{24}),
\ba
&&m=\fr{GM}{a}~~,~~R^2=\tr^2\sin^2\tte~~,~~\tr^2=\Br^2 - a^2, ~~{\rm and}\nn
\\
&&dR^2+dz^2+R^2d\vf^2=\fr{\Br^2 -a^2\cos^2\tte}{r^2 - a^2}dr^2+R^2d\vf^2+(\Br^2- a^2\cos^2\tte) d\tte^2.
\lb{25}
\ea
Both $\tr$ and $\Br$ are constant on prolate spheroids confocal with the shell. $\tte$ is constant on confocal hyperboloids and becomes the spherical polar $\te$ at infinity. In the equatorial region $\tr=R+\OO(z^2/a^2)$. When $b<\tr\ll a$ the potential becomes that of a line of mass per unit length $M/(2a)=m/(2G)$, but at large $\tr$, $\psi\ra GM/\tr$. The surface density of mass on the prolate \sp is
\be
\si=\fr{M}{4\pi ab\sq{\fr{b^2}{a^2}+\sin^2\tte}}.
\lb{26}
\ee
 A static charged prolate spheroidal conductor has an electrical potential of the form (\ref{24}) but with the charge, $q$, replacing mass. If we now freeze the charge density (\ref{26}) onto the \sp by making it an insulator and rotate it about the axis with \av $\Om<c/b$, we find the magnetic field is uniform inside the shell and outside the magneto-static potential $\chi$ is given by
 \be
 \chi=q\Om\fr{b}{a}^2Q_1(\fr{\Br}{a}),
 \lb{27}
 \ee
 $Q_1$ is the Legendre function of the second kind and $\Br/a>1$.
 
 We now consider two confocal prolate \spp with positive charges $q_1$ and $q_2$. Each lies on an equipotential of the other so if they are both conductors their charge distributions is unaltered by the field of the other \sp and then rotate the \spp with \avv $\Om_1>0$ and $\Om_2<0$. The magnetic field is the sum of the fields of each, but as $\Om_2<0$, they tend to cancel, except in the region between the \spp$\!\!$. Externally the magnetic field potential is
 \be
 \chi=\fr{1}{a^2}(q_1\Om_1b_1^2+q_2\Om_2b_2^2)Q_1
\!\!\lll     \fr{\Br}{a}  \rrr\cos\tte
\lb{28}
 \ee
 which is zero if we chose 
 \be
 \Om_2= - \fr{q_1b_1^2}{q_2b_2^2}\Om_1.
 \lb{29}
 \ee
 The magnetic field inside both \spp is of course uniform. We are interested in tall thin \spp with $b_1<b_2\ll a$ and with the above choice the field inside both is
 \be
 B_I=\fr{q_1\Om_1}{a}\lll1-\fr{b_1^2}{b_2^2}\rrr+\OO\lll    \fr{b^2}{a^2}    \rrr\!.
 \lb{210}
 \ee
 In this   thin r\'egime the field between the \spp is approximately uniform in the equatorial region and carries equal and opposite flux to the field inside both so
  \be
B_{II}= - \fr{q_1\Om_1b_1^2}{ab_2^2}.
\lb{211}
 \ee
 The electrical potential outside both is given by (\ref{24}) with $m=(q_1+q_2)/a$.


\sect{Spheroidal shells in \GR}
 We take the metrics in Weyl's form for empty regions
 
 \be
ds^2=e^{-2\psi}(dt+A_idx^i)^2 - e^{2\psi}\LLL  e^{2k}\lll   dz^2+dR^2 \rrr + R^2d\vf^2   \RRR\!.
\lb{31}
\ee
where $k=0$ on the axis and $\bA\!\!=0$ when we deal with statics. The transformation from $z, R$ to $\Br, \tte$ \coo is the same as in flat space considered earlier. Weyl showed that for axially symmetric statics Einstein's \eeee imply $\na^2\psi=0$ where $\na^2$ is the flat space operator. The simplest spheroidal solution is the same as the classical one (\ref{24}) and the corresponding metric function $e^k$  is
\be
e^{k}=\LLL    1+\lll\fr{a}{\tr}\sin\tte\rrr^2    \RRR^{-m^2/2}.
\lb{32}
\ee
This metric can be generated by a single  spheroidal shell the metric inside being flat. Babala \cite{Ba} gives expressions for the stresses needed to support the shell against its own gravity. For prolate shells the energy conditions are most restrictive at the equator and the dominant energy condition is satisfied provided $|p_\te|<\si$ there. From this we find that at fixed mass per unit length $m/2$,  the dominant energy condition  is always {\it violated} if the spheroidal shell is too tall so the cylindrical limit is not attainable. Nevertheless for quite relativistic $m$, axial ratios of order $100$ are attainable without violating the energy conditions so a cylindrical treatment is valid as an approximation in the equatorial region. For a \sp  of semi axes $\sq{a^2+b^2}, b$, Babala's condition [under his \eee (12)] yields
\be
m^2+(X^2-1)(1-Y)+\ha X^2(1-Y^{-1})<m X,
\lb{33}
\ee
where $ X^2=1+b^2/a^2$ and $ Y= [X^2/(X^2-1)]^{m^2/2}$.
For large axial ratios, $X$ must be close to one and the above restriction on the axial ratio becomes
\be
Y<\fr{1}{1-2m(1-m)},
\lb{34}
\ee
corresponding to the restriction on the axial ratio 
\be
\al=\fr{\sq{a^2+b^2}}{b}<\fr{1}{[1-2m(1-m)]^{1/m^2}},
\lb{35}
\ee
$\al$   is the axial ratio as measured in Weyl's \coo which exaggerate elongation. In the internal flat space its axial ratio is less by a factor $Y$ which is about $2$ for the three cases given below:
For $m = 1/3,  \,Y< 9/5,\, \,\al<198$,  for $m = 2/5, \, Y< 25/13, \,\,\al<60$,  and 
for $m = 1/2,  \,Y< 2, \,\,\al<16$. Since parallel matter currents repel, these conditions will be slightly alleviated for a rotating spheroid.

When we have two oppositely rotating spheroidal shells we may choose the rotation of the outer shell to annul the \am of the inner one. That ensures that there is no gravomagnetic moment of the whole system. Just as in electricity it is possible to choose rates within the outer shell so that there is no gravomagnetic field outside. The external field will then be static and predominantly of the form governed by \eeee (\ref{24}), (\ref{28}), (\ref{29}) with $\bA=0$. However there may be some higher moment terms with 
\be
\psi=\psi_0+\sum_{n=1}^\inf a_nP_n(\tmu)Q_n\lll   \fr{\Br}{a}  \rrr
\lb{36}
\ee
as given by Quevedo \cite{Qu}. We are at liberty to choose the space within the inner shell to be flat space in rotating axes, i.e., with a uniform \gm field.  For any chosen form of the inner shell $B_n$ and $\psi$   are continuous while discontinuities in $B_{||}$ and $\bn\cd\!\bna\psi$  give the matter currents and mass density on the shell. Between the shells the \gm field is in the $z$ direction on the equator by symmetry and will be close to that direction in the whole of the long straight region of tall spheroids. The Ernst \eeee in the empty region can be written in terms of flat space operators:
\be
\bna\!\cd\!\!\LLL   e^\psi\bna\!\!\lll    e^{-4\psi} + \chi^2            \rrr                \RRR=0~~~{\rm and}~~~\bna\!\cd\!(e^{4\psi}\bna\chi)=0,
\lb{37}
\ee
and those imply
\be
\na^2\psi=\tha e^{4\psi}|\bna\chi|^2.
\lb{38}
\ee
In regions where the field is along the $z$ direction with $\chi=Hz$, the second of (\ref{37}) is automatically satisfied if $\psi$ is a function of $R$ and the (\ref{38}) may be written
\be
Rd_R(Rd_R\psi)=\tha H^2R^2e^{4\psi}
\lb{39}
\ee
which \eee is readily solved by writing $W=\psi+\tha\ln R$ and using $\ln R$ as the independent variable. We then recover the usual solution for rotating cylinders (see \cite{JM}, \cite{Le}, \cite{ES}):
\be
e^{-2\psi}=\tha HRC^{-1}\LLL     \lll\fr{R}{R_0}\rrr^{-C}\!\! - \lll\fr{R}{R_0}\rrr^C    \RRR\!,
\lb{310}
\ee
where $C$ and $R_0$ are the constants of integration. For the region containing the axis $C=1$.

This derivation of the solution for rotating cylinders as a limiting
case of thin prolate ellipsoids
does not seem to be given before.

 \sect{ Metrics with oppositely rotating cylinders in \GR}

 Consider    two massive cylindrical shells rotating at constant angular velocity in opposite directions around a common axis of symmetry and surrounded on each side by empty space. The empty \sss within the interior shell which we call shell \s$\!$ or simply \s$\!$  
 is flat. It  is   dragged around by the rotations of the shells. We therefore write its metric in cylindrical Minkowski coordinates rotating with constant angular velocity $\B\om$; say, $ \{\B x^\mu\}= \{\B x^0= \B t, \, \B x^1=\B R,\, \B x^2= \B z, \,\B x^3=\vf \}$, then
 \be
 d\B s^2=\B \gmn d\B x^\mu d\B x^\nu=(1-\B v^2)d\B t^2+2\B R\,\B vd\B t d\vf - d\B R^2 - d\B z^2-\B R^2d\vf^2~~~{\rm with}~~~\B v={\B R}{\B\om}~~{\rm for}~~ {\B R}\le\B R_1.
 \lb{41}
 \ee
 $\B R_1$ denotes  the position of \s.
 
 While within the inner cylinder \sss is flat,   seen from there the \sss at infinity rotates at an angular speed $-\B \om$. Taking a global view we say that the flat \sss within the inner cylinder  rotates at a rate $\B\om$. A particle of rest mass $m_0$ which is forced to move in a circle at a rate $- \B\om$ with respect to the inner flat space has momentum
 \be
 \mbox{\bm $\B p$}=\fr{m_0\mbox{\bm $\B v$}}{\sq{ 1-\B v^2}}= - \fr{m_0\B v}{\sq{ 1-\B v^2}}\mbox{\bm $n$}_\vf.
 \lb{42}
 \ee
 Its proper rate of change is 
\be
\fr{d\mbox{\bm $\B p$}}{\sq{ 1-\B v^2}d\B t}= - \fr{m_0\B v}{1-\B v^2}\fr{d\mbox{\bm $n$}_\vf}{d\B t}= - \fr{m_0\B v^2 }{1-\B v^2} \fr{\mbox{\bm $ n$}_R }{\B R}.
\lb{43}
\ee
Thus the centrifugal force $F_c$  induced on a globally static particle is 
\be
F_c=  \fr{m_0\B v^2}{ 1-\B v^2}\fr{1}{\B R}.
\lb{44}
\ee
 
 Between   \s $\!$ and \ss$\!$  the metric is that of Lewis \cite{Le}.  The metric in the form (\ref{31}) offers some difficulties in finding physical  values for the parameters of the second    outer shell, i.e. one that satisfies the dominant energy conditions and does not rotate faster than the velocity of light. The following coordinates are   more   convenient. They  
resemble  those of da Silva {\it et al} \cite{SHSW}.  In coordinates $ \{x^\mu\}= \{x^0=  t, \, x^1=R,\, x^2=  z, \,x^3=\vf \}$, the metric, which contains   4 parameters, reads as follows:
 \be
 ds^2=\gmn dx^\mu dx^\nu= fdt^2+2\fr{h}{\bt}dtd\vf - R^{2(s^2-s)}(dR^2+dz^2)-\fr{l}{\bt^2}d\vf^2~~{\rm for} ~~R_1\le R\le R_2.
 \lb{45}
 \ee
In this,  
\ba
&&f\!=\ep\, R \lll  W-  \fr{(\vp+1)^2}{4W} \rrr,~h\!=\ep\,R\lll  W - \fr{(\vp^2 - 1)}{4W} \rrr,~l\!=\ep\,R\lll \fr{(\vp-1)^2}{4W} - W\rrr\!,\nn
\\
&&~~~~~~~~~~~~~~~~~~~~~~W\!=\!\al R^{2s-1}>0 ~~~{\rm and}~~~\ep =\pm1.
\lb{46}
\ea
The metric is invariant under linear transformations of $t, \vf$ and can be transformed locally to the static Levi-Civita metric \cite{St}. 
The parameter $\bt$ is introduced to normalize the angle $\vf$ to vary from 0 to $2\pi$ in the whole  of $\rm \sss\!\!$; vacuum cylindrical \ssss  with matter sources   have in general conicities\footnote{Conicity can be geometrically defined far away from the axis which may be regular. For instance consider a truncated normal cone without a peak. One starts from a particular circle with radius $r_1$ and  length of the circumference $2\pi r_1$. On moves to another circumference with radius $r_2$ and measure its proper length $2\pi r_2$. One then compares the ratio of these   lengths $r_2/r_1$ with the   distance between the two circles. This is a measure of the conicity.
 It is shown in \cite{BLSZ} that conicity arises generally outside cylinders of perfect fluids.} \cite{BLSZ}. 
Without loss of generality we may assume $\bt>0$.

 Outside shell \ss\,  \sss is empty and static;   the metric is that of Levi-Civita. In coordinates $ \{X^\mu\}= \{X^0=  T, \, X^1=\td R,\, X^2=  Z,\, X^3=\vf \}$ the metric has the form (see, e.g., \cite{ES}):
 \be
 d\td s^2=\td g_{\mn}dX^\mu dX^\nu= \td R^{2m}dT^2 - \td R^{2(m^2-m)}(d\td R^2 + dZ^2) -\fr{1}{\td\bt^2}\td R^{2(1-m)}d\vf^2~~{\rm for}~~\td R\ge \td R_2.
 \lb{47}
 \ee
 The constant $\td\bt>0$ has a role similar to $\bt>0$ and $m$ is analogous to the parameter $m$ defined in (\ref{25}). 
 
 Altogether there  are 6 parameters involved in these metrics: $\al>0, \bt>0, \vp,  s, m$ and $\td\bt>0$    in addition to    the ``radii" of the shells $\B R_1, R_1, R_2$  and $\td R_2$.  $\al$ is a scale factor typical of spacetimes without an intrinsically defined scale. $\bt$ characterizes the conicity of spacetime between the two shells and   $\td\bt$ the conicity of spacetime outside   the outer shell. $s$ is associated with the mass of the inner shell. 
  $\vp$ is the parameter associated with the Coriolis force and the centrifugal force induced by the rotation of the cylinders. We shall refer to $\vp$ as the parameter of induced centrifugal forces.
 
    The metric on  shell \s\,  is obtained from (\ref{41}) in which we set $\B R=\B R_1$ or from (\ref{45}) by    setting $R=R_1$:
\be
ds_1^2=  (1-\B v_1^2)d\B t^2+2\B R_1\,\B v_1d\B t d\vf - d\B z^2-\B R_1^2d\vf^2=f_1dt^2+2\fr{h_1}{\bt}dtd\vf - R_1^{2(s^2-s)}dz^2-\fr{l_1}{\bt^2}d\vf^2.
\lb{48}
\ee
In this $f_1,h_1$ and $l_1$ are the metric components (\ref{46}) with
\be
W~~~\ra~~~W_1=\al R_1^{2s-1}.
\lb{49}
\ee
 $W_1$ is thus a measure of the distance of \s\, to the axis.
  The   two metrics in (\ref{48}) describe the same hypersurface.  We must thus have:
\be
\sqrt{1-\B v_1^2}\B t=\sqrt{f_1}t ~~~{\rm and }~~~\B z=R_1^{s^2-s} z,
\lb{410}
\ee
up to constants of integration. We also need equality of the two remaining terms in the metric (\ref{48}). This implies:
\be
\B R_1=\fr{1}{\bt}\sqrt{l_1}~~~ {\rm and}  ~~~\B v_1=  \ep\lll W_1 - \fr{\vp^2-1}{4W_1} \rrr\!,
\lb{411}
\ee
$\B v_1$ is the fastest ``dragging velocity" of the flat interior.  
 
 The metric of   \ss\,  is obtained from (\ref{45}) in which we set $\td R= \td R_2$ or from  (\ref{47}) in which we set $  R=  R_2$:
\be
ds_2^2= \td R_2^{2m}dT^2 - \td R_2^{2(m^2-m)}dZ^2 -\fr{1}{\td\bt^2}\td R_2^{2(1-m)}d\vf^2=f_2dt^2+2\fr{h_2}{\bt}dtd\vf - R_2^{2(s^2-s)}dz^2-\fr{l_2}{\bt^2}d\vf^2.
\lb{412}
\ee
 The equality implies, among other things, that the term in $dtd\vf$   is absent from $ds^2_2$, i.e.,
\be
h_2=0~~~{\rm or}~~~W_2^2= {\4}(\vp^2 - 1) ~~;~~{\rm so}~~~ f_2=-\ep \fr{R_2}{2W_2} (\vp+1)~~~,~~~l_2=- \ep  \fr{R_2}{2W_2}  (\vp-1).
\lb{413}
\ee
Other junction conditions will be dealt with  below. Since $W^2_2>0$  we see that
\be
\vp^2>1~~~{\rm and}~~~\B v_1=\ep\lll W_1 - \fr{W_2^2}{W_1}\rrr.
\lb{414}
\ee
The  greatest dragging velocity $\B v_1$  of the flat interior near  \s \, depends essentially on the positions of the   shells, for given $\al$ and $s$.   
The three other junction conditions, not yet mentioned, are similar to (\ref{410}) and the first of (\ref{411}):
\be
\sq{f_2}t=\td R_2^mT~~~,~~~R_2^{s^2-s}z=\td R_2^{m^2-m}Z~~~{\rm and}~~~\fr{1}{\bt}\sq{l_2}=\fr{1}{\td\bt}\td R_2^{1-m}.
\lb{415}
\ee
 
A word about  the conditions that   \sss between \s\, and \ss be locally Minkowski.
These conditions are necessary and amount to ask that  $g_{00}>0$ and $g_{33}<0$. If we add the condition that \ss\, be outside of \s, we have altogether three inequalities that must be satisfied\footnote{We do not consider the possibility of $g_{00}<0, \, g_{33}>0$ when the role of $t$ and $\vf$ would be interchanged.}:
\be
f>0~~~,~~~l>0~~~,~~~{\rm and}~~~\fr{R_2}{R_1}>1.
\lb{416}
\ee
This translates into the following conditions on the parameters. If
\ba
  s>\ha~,&&\vp>1~~,~~\ep =-1 ~~~{\rm and}~~~ \4(\vp-1)^2<W_1^2<W_2^2<\4(\vp+1)^2, 
\lb{417}
 \\{\rm or}&&\vp< - 1~~,~~\ep =+1~~~{\rm and}~~~ \4(\vp+1)^2<W_1^2<W_2^2<\4(\vp-1)^2, 
\lb{418}
\ea
but if
\ba
  s<\ha~,&&\vp>1~~,~~\ep=-1 ~~~{\rm and}~~~ \4(\vp-1)^2<W_2^2<W_1^2<\4(\vp+1)^2,
\lb{419}
 \\{\rm or}&&\vp< - 1~~,~~\ep =+1 ~~~{\rm and}~~~ \4(\vp+1)^2<W_2^2<W_1^2<\4(\vp-1)^2 .
\lb{420}
\ea
In either of these cases $\ep\vp<0$ and one can easily check that the greatest dragging velocity never exceeds the speed of light:
\be
- 1<\B v_1<1.
\lb{421}
\ee

We now turn our attention to the structure of the shells.


\sect{ Energy densities and pressures in the shells }
The two shells have   similar metrics (\ref{48}) and (\ref{412}). We write them collectively in the same   form without   indices 1 or 2:
\be
d\si^2=\ga_{ab}dx^adx^b=fdt^2+2\fr{h}{\bt}dtd\vf - R^{2(s^2-s)}dz^2-\fr{l}{\bt^2}d\vf^2~~{\rm with}~~\{x^a\}=\{x^0=t, x^2=z, x^3=\vf\}.
\lb{51}
\ee
This is the metric of a \hy $R=$const in a \sss whose metric $\gmn$ is given by (\ref{45}).
We assume   the shells to be     in the form of   two-dimensional fluids rotating with angular velocity $\Om$. The three velocity components $u^a$ are thus $ u^0, u^2=0, u^3=u^0\Om$,  with $u^0$ defined by the usual normalization $\ga_{ab}u^au^b=u_au^a=1$. Let $\si$ be the mass-energy density, $p_\vf$ the pressure or tension in the loops,  and $p_z$ the vertical pressure or tension. The energy-momentum  tensor of such a flow $\tau^b_a$ is necessarily of the following form in which all the components are constants:
\be
\tau_0^0=(\si+p_\vf)u^0u_0 - p_\vf ~,~\tau^0_3=(\si+p_\vf)u^0u_3 ~,~\tau_3^3=(\si+p_\vf)u^3u_3 - p_\vf~{\rm and}~\tau^2_2= - p_z.
\lb{52}
\ee 
There is also a $\tau^3_0$ component similar to $\tau^0_3$; other components are equal to zero. From these expressions and with $u_au^a=1$ we may calculate the relevant physical quantities $\si, p_\vf, p_z$ and $\Om$. Set
\be
\Te^2= (\tau^0_0 - \tau^3_3)^2+4\tau^0_3\tau^3_0.
\lb{53}
\ee
Then,
\be
\si=\tha\![(\tau^0_0+\tau^3_3)+\Te]~~,~~p_\vf=\tha\!\LLL - (\tau^0_0+\tau^3_3)+\Te\RRR~~~,~~~p_z= - \tau^2_2,
\lb{54}
\ee
and
\be
\Om=\fr{d\vf}{dt}=  \fr{1}{2\tau^0_3} \LLL - (\tau^0_0 - \tau^3_3)+\Te\RRR~~~\Ra~~~v=\fr{\sq{l}}{\bt\sq{f}}\Om,
\lb{55}
\ee
  $v$ is the proper velocity of the shell.

Next we can easily calculate the external curvature tensor  components from both sides of the shell, say $K_{ab}$ and $\B K_{ab}$. The surface \em tensor\footnote{ In Israel formalism \cite{Is} unit normal vectors to the shell have the same orientation and $\pm L_{ab}=K_{ab} -\B K_{ab}$; the sign depends on the orientation of the  normals. In \cite{GK} the unit normal vectors are oriented in their own \sss and $L_{ab}=K_{ab} +\B K_{ab}$. This convention which is adopted here is less ambiguous when, for instance, the \sss is closed on both sides of the shell; it is also more symmetrical.} $\tau^b_a $   is given by
  \be
  \tau^b_a=\1k(\de^b_aL^c_c - L^b_a)~~~{\rm where}~~~L^b_a=K^b_a+\B K^b_a~~~{\rm with}~~~\ka=\fr{8\pi G}{c^4}.
  \lb{56}
  \ee
If $n_\mu=\gmn n^\nu$ is the normal to the shell     the expression of the external curvature components say $K_{ab}$ (and a similar expression for $\B K_{ab}$) is as follows
\be
K_{ab}= - \fr{\di x^\mu}{\di x^a}\fr{\di x^\nu}{\di x^b}D_\mu n_\nu,
\lb{57}
\ee
$D_\mu$ is a four covariant derivative  in terms of the \sss metric $\gmn$   (or $\B\gmn$ or  $\td g_{\mn}$). 
For cylindrical shells and in the coordinates adopted,  $K_{ab}$ is particularly simple to calculate:
\be
K_{ab}= - \tha n^1\di_1 g_{ab}.
\lb{58}
\ee
  
With the tensors $K^b_a$  and $\B K^b_a$ we construct   the tensors $L^b_a$  and $\B L^b_a$ and with them the \em tensor of the shell given     in (\ref{66}).

So much about generalities.

\sect{ Example of two shells of dust }
 
 In such shells there is no pressure in either direction $z$ or $\vf$.  If
 \be
 p_z|_1=p_z|_2=0,
 \lb{61}
 \ee
it follows from
(\ref{52}) and the evaluation of $\tau^2_2=0$, that
\be
\fr{\bt}{\sq{l_1}}=\fr{1}{R_1^\Si}~~~{\rm and}~~~\fr{1}{\td R_2^M}=\fr{1}{R_2^\Si}.
\lb{62}
\ee
The first equality determines $\bt$ and the second equality $\td\bt$ through $\fr{1}{\bt}\sq{l_2}=\fr{1}{\td\bt}\td R_2^{1-m}$, see (\ref{415}).  
 If in addition
 \be
 p_\vf|_1=p_\vf|_2=0,
 \lb{63}
 \ee
 it follows from (\ref{54}) and evaluating $\tau^0_0, \tau^3_3, \tau^3_0$ and $\tau^0_3$ that
 \be
p_\vf|_1=0~~\Ra~~ \Te|_1=\fr{1}{\ka R_1^\Si}2s(1-s)~~~{\rm and}~~~p_\vf|_2=0~~\Ra~~ \Te|_2= \fr{1}{\ka R_2^\Si}[2m(1-m) - 2s(1-s)].
 \lb{64}
 \ee
 The energy per unit length and the velocities of the shells reduce thus  to
 \be
 \si|_1=\fr{1}{\ka R_1^\Si}2s(1-s) ~~~,~~~v_1=  \fr{ \ep_\vp}{|\vp-1|}\Bigg\{ \fr{\LLL (\vp -1)/2W_1  \RRR^2-1 }{ 1-\LLL (\vp+1)/2 W_1 \RRR^2}      \Bigg\}^{1/2} \!\!\!\!\! \LLL   (2s-1)\vp - 1 + 2s(1-s)\RRR,
\lb{65}
 \ee
 and
 \ba
 &&\si|_2=\fr{1}{\ka R_2^\Si}[2m(1-m) - 2s(1-s)],
 \lb{66}\\~
 &&v_2=  - \fr{\ep_\vp}{\sq{\vp^2 - 1}}\LLL (1-2s)\vp + 1 - 2m +2m(1-m) - 2s(1-s) \RRR\!.
 \lb{67}
  \ea
  
  The energy condition
\be
\si|_1>0~~~{\rm amounts~to}~~~0<s<1,
\lb{68}
\ee
 but if we add the condition  that $\si|_2>0$,
 \be
{\rm then~~ either}~~ \ha<s<1~~, ~~1-s<m<s~~~~~{\rm or}~~~~~0<s<\ha~~, ~~s<m<1-s.
 \lb{69}
 \ee
 
 The parameters in these metrics and the associated physical quantities are intertwined in complicated ways. We can however see in (\ref{65}) that $s$ characterizes the energy  per unit length of the inner cylinder. $W_1$ for a given energy per unit length is a measure of the radius of the inner shell as we  noticed before. We also noticed that $m$ represents the mass per unit length of spacetime far from the cylinders in the $R$ direction.

Equations (\ref{61})  and (\ref{63}) are 2 polynomials of order 2 in $\vp(s)$ and $\vp(s, m)$, see (\ref{64}) and (\ref{66}). There are thus twice 2 roots, say $\vp_\pm(s)$ and $\vp_\pm(s,m)$, which must be equal. This gives 4 possible solutions  for $m(s)$ or $s(m)$.  Mathematica solves such equations with great facility.
 It shows that among the four possible solutions   only one   satisfies the energy conditions  in which 
 \be
 0<s<\tha~~~{\rm  with} ~~~s<m<1-s.
 \lb{610}
 \ee
 van Stockum \cite{vS} constructed in 1937 a rotating cylindrical shell of dust and showed that this is only possible if $s<\tha$.

\begin{figure}[ht]
   \centering
   \includegraphics[width=10 cm]{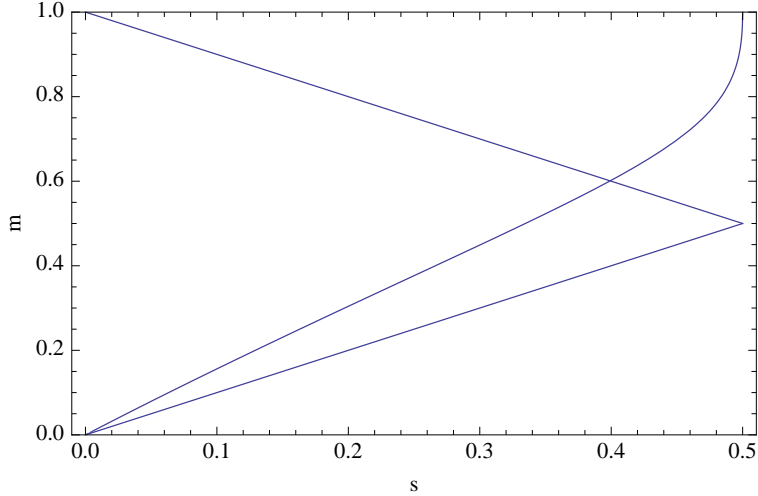} 
   \caption{\small  The function $m(s)$, implied by conditions (\ref{64}), relates the total mass per unit length, $m$,  to the parameter characterizing the mass per unit length of the inner shell, $s$. The straight lines are the limits $m=s$ and $m=1-s$ imposed by the energy condition on \ss$\!$. Above the triangle, the energy density of the outer shell is negative.}
\end{figure}   
Figure 1 represents $m(s)$.  From  this figure we can see   that the range of values which satisfy the energy conditions are  in fact
\be
0\le s\les 0.4~~~{\rm and}~~~0\le m(s)\les 0.6.
\lb{611}
\ee

 We also find  that $\vp< - 1$ which implies, see (\ref{413}), $\ep=1$ and, within the limits  of $s$, that is 
\be
0\le s\les0.4~~~\Ra~~~ -1.5\les\vp\le-1.
\lb{612}
\ee

 Quantities analyzed so far depend on one parameter $s$ associated with the mass of the inner shell. However, the inner shell ``radius", or better $W_1=\al/R_1^{1-2s}$, is not   fixed. According to (\ref{420}):
\be
W_{1min}=W_2=\sq{\fr{\vp^2-1}{4}}<W_1<W_{1max}=\fr{1-\vp}{2}.
\lb{613}  
\ee

It is useful to remember expression  (\ref{43}) from which follows that there is a smallest ``radius" $R_{1min}$:
 \be
 W_{1max}=\fr{\al}{R^{1-2s}_{1min}}.
 \lb{614}
 \ee
 When $W_1\ra W_{1max}$ the following happens: Since $\vp<-1$,  $\ep =1$ and the 
 {\it proper radius} of the inner shell  tends to zero, see (\ref{411}):
\be
 \B R_1=\fr{1}{\bt}\sq{l}=\fr{1}{\bt}\LLL   R_1\lll \fr{W_{1max}^2}{W_1} - W_1\rrr \RRR^{1/2}~\ra~0.
 \lb{615}
 \ee 
 As $W_1$ approaches its (unattainable) maximal value $W_{1max}$, the metric component $g_{33}=- l/\bt^2\ra0$, the coordinate system becomes unphysical and the proper velocity of the inner shell, see (\ref{613}), tends to zero:
\be
v_1 \propto \LLL  \fr{( W_{1max}/W_1)^2-1}{1 - [(\vp+1)/2W_1]^2}          \RRR^{1/2}~\ra~0.
\lb{616}
\ee
The velocity of the inner shell as seen from the flat space inside approaches that of light and the angular velocity increases without bound as the radius $\B R_1\ra0$.
Calculating  the dragging velocity from (\ref{411}), we indeed find that it tends to the velocity of light:
\be
\B v_1= W_{1max} - \fr{W_{1min}^2}{W_{1max}}=1.
\lb{617}
\ee
 
When, on the contrary,  $W_1=W_{1min}=W_2$, we are dealing with two counter-rotating shells of dust with different energies   per unit length and different velocities whose  total angular momentum is equal to zero and there is no dragging inside.

 For small mass-energies per unit length of the shells, i.e. in the Newtonian limit, $s\ll1$, and for
\be
m \sim 1.618 s~~~,~~~\vp\sim-1 - 4s^3~~~, ~~~W_2 \sim \sq{2}s^{3/2}~~~
, ~~~v_2 \sim-1.14 s^{1/2},
\lb{618}
\ee
and 
\be
\fr{W_2}{W_1}=\lll  \fr{R_1}{R_2} \rrr^{1-2s}\sim  \fr{R_1}{R_2}, 
\lb{619}
\ee
we see from (\ref{65}),
\be
v_1\sim s^2\LLL  \fr{1}{2s^3}  \lll  \fr{R_1}{R_2} \rrr^2 -   2s^3   \lll  \fr{R_2}{R_1} \rrr^2   \RRR^{1/2},
\lb{620}
\ee
and,
\be
\fr{|v_2|}{v_1}\sim1.618 \LLL   \lll \fr{R_1}{R_2} \rrr^2  - 4s^6  \lll\fr{R_2}{R_1} \rrr^2  \RRR^{-1/2}~~~{\rm with}~~~\sq{2}s^{3/2}\les \fr{R_1}{R_2} \les 1.
\lb{621}
\ee
We thus see that to have strong dragging, in the classical limit,  $|v_2|/v_1\sim 1.618$,    we need $R_1/R_2 \thickapprox 1$. Otherwise, to have strong dragging we need $R_1/R_2 \thickapprox s^{3/2}$. $R_1/R_2=0.1$ is already very relativistic.  

It is interesting to have some idea of what the ratio of the velocities is in this relativistic case. Figure 2 shows the value of the maximum dragging velocity as a function of the parameter $s$.
\begin{figure}[ht]
   \centering
   \includegraphics[width=10 cm]{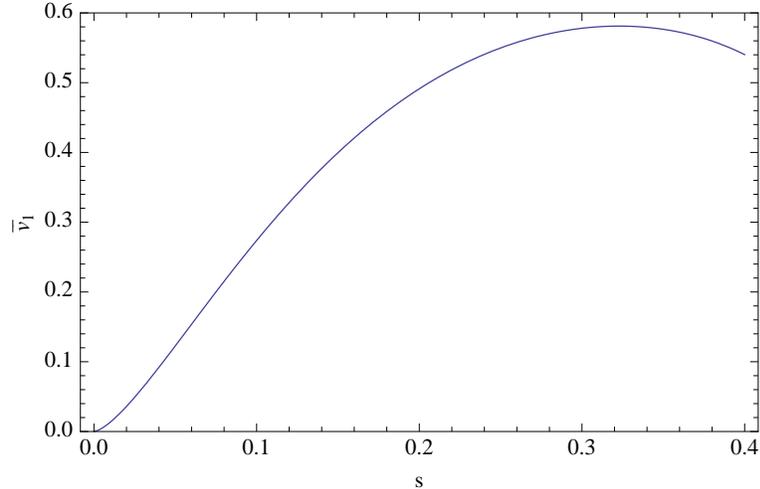} 
   \caption{\small  The dragging velocities $\B v_1$  as functions of  the  parameter of the inner shell energy per unit length $s$ with $R_1/R_2=0.1$. At smaller ratios the maximum would still be higher. }
   \end{figure}
  We notice that the dragging $\B v_1$ never exceeds $0.6$ and is only slightly greater than $(-v_2)$ in the extreme relativistic case when $s\ra 0.4$.  In the Newtonian limit for small velocities,
  \be
  \fr{\B v_1}{(-v_2)}\sim 9.35 v_2^2.
  \lb{622}
  \ee
  
  The velocity of the inner shell never exceeds $0.25$ and the ratio of velocities of the shells $v_1/(-v_2)$  never exceeds $0.5$; in the Newtonian limit
  \be
  \fr{v_1}{v_2}\sim0.124+0.256 v_2^2.
  \lb{623}
  \ee
  {\bf  Acknowledgments}
 \vs
  J.B. and J.K acknowledge the hospitality of the Institute of Astronomy in Cambridge. J.B. acknowledges    the hospitality of the Einstein Institute in Golm, the partial support from the Grant GA\v CR 202/09/00772 of the Czech Republic and the Grant No LC06014 and  MSM0021620860 of the Ministry of Education.
  \vs

  \end{document}